\documentclass[twocolumn,pra,aps,longbibliography,superscriptaddress]{revtex4-2}
\usepackage{amssymb,amsmath,amsfonts}

\usepackage[latin9]{inputenc}
\usepackage{color}
\usepackage{amstext}
\usepackage{enumitem}
\usepackage{graphicx,bm,palatino}
\usepackage[colorlinks=true,linkcolor=blue,urlcolor=blue,citecolor=blue,pdfusetitle]{hyperref}

\usepackage[sc]{mathpazo} 

\usepackage{hyperref,cleveref}
\usepackage[dvipsnames]{xcolor}
\usepackage[caption=false]{subfig}
\makeatletter

\newcommand\ket[1]{\left|#1\right\rangle}
\newcommand\bra[1]{\left\langle #1 \right|}

\usepackage{times}
\usepackage{bbm}

\makeatother

\begin{document}
\title{A Reinforcement Learning Approach to the Design of Quantum Chains for Optimal Energy Transfer}

\author{S. Sgroi}
\affiliation{Centre for Quantum Materials and Technologies, School of Mathematics and Physics, Queen's University Belfast, BT7 1NN, United Kingdom}
\author{G. Zicari}
\affiliation{Centre for Quantum Materials and Technologies, School of Mathematics and Physics, Queen's University Belfast, BT7 1NN, United Kingdom}
\author{A. Imparato}
\affiliation{Department of Physics and Astronomy, University of Aarhus,
Ny Munkegade, Building 1520, DK-8000 Aarhus C, Denmark}
\affiliation{Centre for Quantum Materials and Technologies, School of Mathematics and Physics, Queen's University Belfast, BT7 1NN, United Kingdom}
\author{M. Paternostro}
\affiliation{Universit\`a degli Studi di Palermo, Dipartimento di Fisica e Chimica - Emilio Segr\`e, via Archirafi 36, I-90123 Palermo, Italy}
\affiliation{Centre for Quantum Materials and Technologies, School of Mathematics and Physics, Queen's University Belfast, BT7 1NN, United Kingdom}

\begin{abstract}
We propose a bottom-up approach, based on Reinforcement Learning, to the design of a  chain achieving efficient excitation-transfer performances. We assume distance-dependent interactions among particles arranged in a chain under tight-binding conditions. Starting from two particles and a localised excitation, we gradually increase the number of constitutents of the system so as to improve the transfer probability. We formulate the problem of finding the optimal locations and numbers of particles as a Markov Decision Process: we use Proximal Policy Optimization to find the optimal chain-building policies and the optimal chain configurations under different scenarios. We consider both the case in which the target is a sink connected to the end of the chain and the case in which the target is the right-most particle in the chain. We address the problem of disorder in the chain induced by particle positioning errors. We are able to achieve extremely high excitation transfer in all cases, with different chain configurations and properties depending on the specific conditions.
\end{abstract}

\maketitle

\section{Introduction}
\label{sec:intro}
Studying and optimizing energy, information or state transfer across physical systems are problems of great importance in a multitude of contexts in physics: the field of quantum physics, and quantum technologies in particular, is certainly not an exception. Among the various quantum systems whose transport properties are of special interest, particle chains play a special role for quantum technologies, both for their relative simplicity and for their wide range of applicability. Quantum communications \cite{Gisin2007, Chen_2021} would obviously benefit from a better understanding of transport properties of particle chains together with better tools to design optimal state transfer across them. This would also be beneficial for quantum computing as, for example, particle chains can be used to describe spin-like systems which might be useful to connect distinct quantum processors and registers. For these reasons, various techniques have been developed to realize transport across these structures \cite{SBose2007}.

Designing optimal couplings among the particles in such chains would allows us to avoid or minimize the control we have to exert during the system dynamics to achieve effective transfer. While arbitrary couplings engineering between particles in a chain can be a difficult task to accomplish for generic physical systems, some platforms where the couplings can be distance-dependent, such as ion traps \cite{Harlander2011-zi}, could allow some degree of control over their design.

One possible way to optimizing such couplings is by making use of Reinforcement Learning (RL) \cite{Sutton1998}. Machine Learning techniques have been extensively used in recent years to solve different physical problems with great success \cite{RevModPhys.91.045002}, even in the quantum realm \cite{PhysRevA.107.010101}. In particular, RL has been proved especially useful in the context of quantum control, in some cases even clearly outperforming most commonly used optimal control algorithms \cite{Zhang2019}. RL has also been applied to realise fast transfer across particle chains via magnetic fields control \cite{PhysRevA.97.052333}. However, its potential for optimal quantum system design remains largely unexplored.

In this work, we use a RL approach to design optimal particle chains for excitation transfer when the particles interactions depend on their relative position in space. We consider one excitation at maximum in the chain to ease numerical simulations, and we assume dipole-dipole interactions between particles. However, the approach can be easily extended beyond these conditions, and can, in principle, be directly used in an experimental setting without the need of simulating the system dynamics.

Our approach offers multiple advantages compared to most analytical or numerical optimization approaches. In particular, we deploy a spatial approach to find the optimal chain design instead of considering arbitrary couplings, making it closer to realistic physical problems. We consider the chain as a fully connected quantum network with distance dependent couplings, hence we do not resort to nearest neighbours interactions or other approximations. Furhermore, we allow for a variable number of particles in the chain (instead of fixing it beforehand), while encouraging the use of less nodes if possible.

The paper is organized as follows. In \Cref{sec:problem} we describe the system of interest and introduce physical problem. In particular, we propose a spatial, bottom-up approach to build a particle chain for optimal excitation transfer, which -- in order to address it with RL -- we formulate as a decision process. In \Cref{sec:RL}, we briefly introduce the RL framework and we present our RL approach for optimal chain design. In \Cref{sec:Results} we show our numerical study on the effectiveness of such RL approach, along with our optimal chain solutions under different conditions. In particular, we consider the scenario where we are not interested in coherence preservation in \Cref{sec:R_opt}; we then study the effects of errors and disorder in \Cref{sec:R_err}, where we also introduce a possible adaptation of the original technique to further improve the transfer in this case. Finally, we address the problem of excitation transfer without coherence loss in \Cref{sec:R_uni}. Conclusions are discussed in Section \ref{sec:conclusions}, together with future outlooks.

\section{Physical problem}
\label{sec:problem}
Let us consider a system of two particles, $A$ and $B$, coupled through the Hamiltonian  
\begin{equation}
\label{eq:H_AB}
\hat{H}_{AB} = \Delta E\sum_{i=A,B}\ket{i}\! \bra{i} + V_{AB}(\ket{A}\! \bra{B} + \ket{A}\! \bra{B}) 
\end{equation}
with $\ket{i}$ identifying a state where particle $i=A,B$ is in the excited state. We have assumed that the coupling originates from a dipole-dipole-like interaction whose strength scales with the (dimensionless) distance $d_{AB}$ between them as
\begin{equation}\label{eq:interaction2q}
    V_{AB} = {J}/{d_{AB}^3},
\end{equation}
where the coupling constant $J$ is written in units such that $V_{AB}$ has the dimensions of an energy. This implies that $d_{AB}$ is rescaled by a typical distance dictated by the specifics of the implementation of the chain at hand. 
Note that we assume that both particles have the same 
energy $\Delta E$ and, for the sake of simplicity, we focus on the case in which one excitation at a time is allowed in the whole system.


Let us suppose that $A$ is prepared in the excited state, while $B$ is initially in its ground state. The unitary evolution of the system is governed by \Cref{eq:H_AB}, while we allow for the incoherent transfer of the excitation from $B$ to a sink $S$ through an incoherent damping mechanism. Such evolution is described by a master equation in the Lindblad form
\begin{equation}
\label{eq:ME}
\dot{\hat{\rho}} = - i [\hat{H}_{AB}, \hat{\rho}] + D[\hat{L}] \hat{\rho} \, ,
\end{equation}
where the first term accounts for the unitary evolution while the second  involves the dissipator accounting for the incoherent process. The latter is given by $
D[\hat{L}] \hat{\rho} \equiv\hat{L} \hat{\rho} \hat{L}^\dagger -\left \{\hat{L}^\dagger \hat{L}, \hat{\rho} \right \}/2 $, with the Lindblad operator $\hat{L} \equiv \hat{L}_{sink} = \sqrt{\Gamma}|S\rangle \! \langle B|$ and the damping rate $\Gamma$.

\begin{figure}[b]
	{\bf (a)}
	\centering\includegraphics[width=\columnwidth]{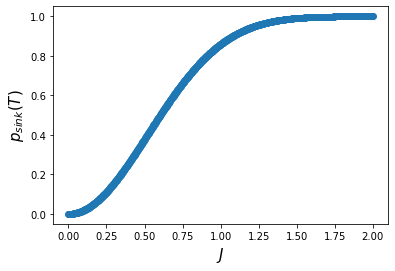}\\
	{\bf (b)}
	\includegraphics[width=\columnwidth]{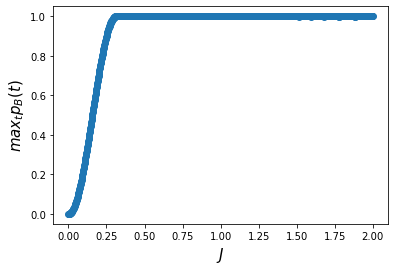}
	\caption{{\bf (a)} Instance of final excitation transferred to the sink [{\bf (b)} Maximum excitation transferred to B] within the evolution time $T$, against the coupling constant $J$ (in units of $\Delta E$) in the case of two particles when the evolution is described by Eq. (\ref{eq:H_AB}). Here $T =5/\Delta E$, $d_{AB}=1$ and $\Gamma_{sink}/\Delta E = 5$. }\label{Figure:2sitesTransfer}
\end{figure}

Given the distance $d_{AB}$, the amount of excitation $p_{sink}(T)$ transferred from $A$ to the sink within a given interval of time $T$  will be determined by the coupling constant $J$: the stronger the coupling between the two particles, the higher will be the population transferred to the sink. A similar behaviour is observed either when one extends the evolution time $T$ while keeping $J$ and $d_{AB}$ constant, or decreases the distance $d_{AB}$ while having $T$ and $J$ fixed. Alternatively, one can consider a second scenario where we are simply interested in the population transfer from $A$ to $B$. In such a case we do not need to include the sink $S$, therefore the system's dynamics is fully unitary, and in \Cref{eq:ME} only the fist term on the right-hand side appears. 
In this case there is a coherent excitation exchange between the two sites, which results in revivals. Regardless of the formulation being chosen, it is natural to cast the problem in terms of maximum population transferred to $B$ within the time interval $T$, i.e. $\max_{t\in [0,T]} p_B(t)$. With such a formulation of the problem, $\max_{t\in [0,T]} p_B(t)$ showcases a monotonic behavior against $J$  [cf. \Cref{Figure:2sitesTransfer}]. However, estimating $\max_{t\in [0,T]} p_B(t)$ is computationally more demanding than calculating a sink population at the end of the evolution, as it requires to track the entire dynamics. Therefore, for the sake of simplicity, we will focus on the first scenario for most of this study, while the second scenario will be addressed in \Cref{sec:R_uni}.

In both cases, given $A$ and $B$, our goal is to enhance the mutual transfer of excitations by designing a suitable particle chain. For simplicity, we assume the particles to have all the same $\Delta E$. Hence, the system Hamiltonian is a straightforward generalisation of \Cref{eq:H_AB}, i.e.
\begin{align}
\label{eq:H_chain}
\hat{H} = \Delta E\sum_{i}\ket{i}\! \bra{i} + \sum_{j \ne i} V_{ij}(\ket{i}\! \bra{j} + \ket{j}\! \bra{i}) \, ,
\end{align}
where the sum runs over all the possible particles of the chain, while the hopping potential is
\begin{equation}
    V_{ij} = \frac{J}{|x_i-x_j|^3}.
\end{equation}
The coupling constant $J$ is assumed to be equal across the chain, while  $x_i$ and $x_j$ denote the positions of the $i$-th and the $j$-th particle, respectively. {The Hamiltonian in \Cref{eq:H_chain}, which straightforwardly generalizes the model in \Cref{eq:H_AB}, represents a tight binding-model where the states $\{\ket{i}\}$ are associated with some spatial degrees of freedom, i.e. the system is found in the state $\ket{i}$ when the excitation is located in the particle in site $i$ while $\Delta E$ is the site energy resulting from the presence of such excitation}. We stress that, in our model, all particles interact with each other, making the chain a fully connected quantum network. 

\begin{figure}
\centering\includegraphics[width=0.8\columnwidth]{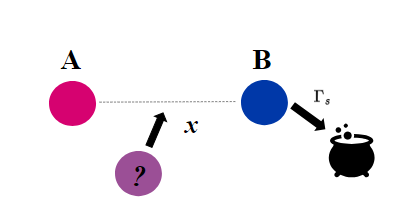}
	\caption{Sketch of the physical problem: two mutually interacting particles, A and B, undergo excitation-transfer processes from A, initially prepared in its excited state, to B. The latter is also incoherently coupled to a sink. We consider this as a starting configuration for the building up of a network: one by one, we add particles between A and B and seek for the potential increase of the excitation-transfer efficiency. The inter-particle interactions depends on their relative positions.}\label{Figure:transfer}
\end{figure}

Designing an optimal chain is equivalent to finding the best number and relative positions of its elements. To achieve this goal, we propose a bottom-up approach: we start with a chain composed of $A$ and $B$ only. Then, for a certain number of steps, we decide if and where to add individual sites to the chain and see how the population transfer is improved [cf. \Cref{Figure:transfer}]. We are interested in the cumulative improvement, i.e. in the final or maximal population transfer accomplished at the very end of the building process.

Having formulated the problem of chain design as a decision process, the next natural step is the search of its optimal working point through RL.

\section{Reinforcement Learning approach}
\label{sec:RL}
RL problems are characterized by an agent observing and interacting with its environment while being assigned with a specific task. The performance of the agent with respect to the given task must be expressible via a numerical feedback, called \emph{reward}, received as results of its interactions with the environment. The purpose of the agent is to learn how to interact optimally with the environment by trial and error, trying to maximize its (long term) reward.

\begin{figure*}[t]
{\bf (a)}\hskip7cm{\bf (b)}\\
\centering\includegraphics[width=\columnwidth]{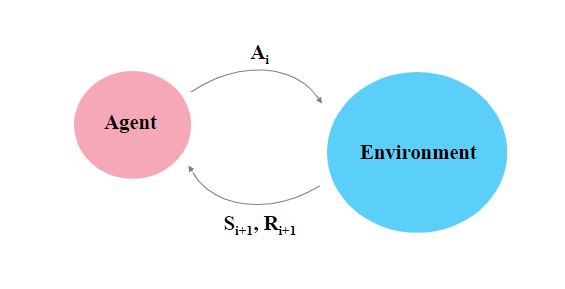}\centering\includegraphics[width=\columnwidth]{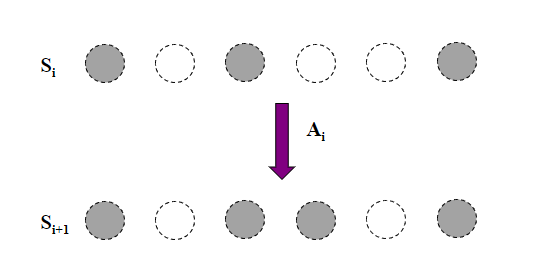}\\
	\caption{{\bf (a)} Sketch of a Markov Decision Process. An Agent interacts with its environment with the purpose of performing a given task. At each interaction step, the Agent observes the state of the environment $S_i$ and, based on this observation, performs an action $A_i$, which in turn changes the state of the environment and hence the next observation $S_{i+1}$. Based on its performance relative to the task, the agent receives feedback in the form of a reward $R_{i+1}$. {\bf (b)} Agent's operations in the discretized-space configuration. The available spacial cells are illustrated by the dotted circles and the presence of a particle is represented by a filled area. Following observation $S_i$, the Agent's action is to decide which cell (if any) to populate at step $i+1$.}\label{Figure:RL}
\end{figure*}

Such agent-environment, interactions-feedback process can be formalized as a Markov decision process (MDP) \cite{Sutton1998}: at each interaction step, the agent observe the state of the environment $S_i$ and performs an action $A_i$ based on the current observation. As a result, the environment state is changed (the next observation will be $S_{i+1}$) and the agent receives a reward $R_{i+1}$ [cf. \Cref{Figure:RL} {\bf (a)}]. Here we consider an episodic task in which the environment is reset after a certain number of interactions, an episode, or after reaching a terminal state.

The  behaviour of the agent can be expressed by the {\it agent's policy} $\pi(A_i=a|S_i=s)$, which is the probability of performing an action $a$ at step $i$ conditioned on the observation of state $s$. The agent's goal will be to find the policy $\pi_{opt}(A_i=a|S_i=s)$ that maximizes the {\it return function}
\begin{equation}
    G_i = R_{i+1} + \gamma R_{i+2} + \gamma^2 R_{i+3} + \dots \, ,
\end{equation}
where the discount factor $\gamma \in [0,1]$ ($\gamma=1$ being included only in episodic tasks) expresses how much we want to weight immediate and long term rewards. 

RL provides numerous ways of approaching such problems. One possibility consists in parametrizing the policy $\pi_{\theta}(A_i=a|S_i=s)$, usually via a Neural Network, and optimizing the expectation value of $G$, or an estimate of it, with respect to the parameters $\theta$, e.g. via gradient ascent. To perform such optimization, data needs to be gathered by observing the MDP. This can also be done in various ways and multiple algorithms have been proposed. In our work we used Proximal Policy Optimization (PPO) \cite{schulman2017proximal}, which is one of the most successful and widespread algorithms today.

Regardless of the particular algorithm being chosen, a RL approach to our problem requires the definition of the corresponding MDP, i.e. to define Observations, Actions and Rewards. This can be done straightforwardly in our case: at each step, the Observation will be the spatial configuration of the chain, i.e. the relative positions of the particles; the Action will be the absence/presence, and the position, of the next particle to be added to the system; the Reward will be the change in the sink population at time $T$, $p_{sink}(T)^{i+1} - p_{sink}(T)^{i}$. Alternatively, in the unitary case, the latter is given by the change in the maximum probability to occupy site $B$ within the time interval $T$, i.e. $\max_{t \in [0,T]} p_B(t)^{i+1} - \max_{ t\in [0,T]} p_B(t)^{i}$.

To describe the physical positions of the particles, we discretize the space between $A$ and $B$ by considering $N_{cells}$ equally spaced cells, all of the same width, so that the state of the environment is described by a binary string of $0$s and $1$s describing the absence or presence of a particle in the respective cell. The Action will be the index of the next cell to populate with a particle, while no particle will be added if a cell is already populated [cf. Figure~\ref{Figure:RL} {\bf (b)} for an illustration of the process]. Although such discretization is not strictly required, its use resulted in a better performance of the preliminary numerical experiments that we have performed and, in general, allows us to provide a simplified description of the approach we have taken.

We introduce an upper bound $\nu_{steps}$ to the number of steps before ending an episode. Needless to say, this sets a constraint to the maximum number of particles that could be allocated in the chain. We set a second strong constraint by imposing the end of an episode whenever the population being transferred from $A$ to $B$ exceeds $0.99$. Both conditions serve the purpose of limiting the physical resources used to build the chain, and can be modified or removed altogether, should it be needed.

\begin{figure}[b]
\centering\includegraphics[width=\columnwidth]{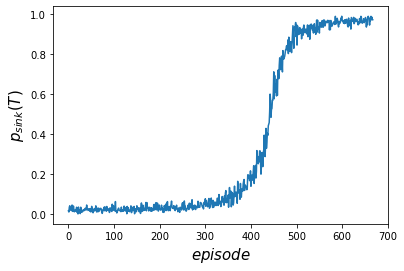}
	\caption{Example of learning process. Notice that $G_0\approx p_{sink}(T)$ if $R_{i+1} = p_{sink}(T)^{i+1} - p_{sink}(T)^{i}$ and $\gamma\approx 1$. Here $d_{AB}=1$, $T \Delta E=5$, $\Gamma_{sink}/\Delta E = 5$, $J/\Delta E=0.05$, while $\gamma/\Delta E=0.99$.}\label{Figure:learning}
\end{figure}

Figure \ref{Figure:learning} shows an example of learning curve for our problem, where it can be seen how the Agent performance, and thus the chain's ability to transfer the excitation, improves episode by episode as a result of the correspondingly improving Agent's policy.
It is worth noticing that, iif no sources of errors, disorder or noise affect the dynamics of the chain, we often would not need to find a fine solution for the RL problem, as in this case we are not interested in the actual policy but in the best chain configuration. Such {\it optimum} will be found during the learning process, before an optimal policy is found.


\section{Case Studies and Results}
\label{sec:Results}
In this Section, we present some numerical results and optimal solutions, i.e. optimal chain designs under various conditions. Unless otherwise specified, we have set $d_{AB}=1$ (in arbitrary units of length), $T \Delta E=5$, $\Gamma_{sink}/\Delta E = 5$, $J/\Delta E=0.05$. For such a  choice of  physical parameters, which allow for an effective illustration of the performance of our protocol,  the excitation transfer is close to zero when the system reduces to a chain only made of particles $A$ and $B$. Quantitatively, we find $p_{sink}(T)^0\approx0.005$ when the sink is present, while $\max_t p_B(t)^{0}\approx0.06$ when the system dynamics if fully unitary. These results were obtained by discretizing the space in $N_{cells}=21$ cells, and setting a maximum of $\nu_{steps}=11$ steps. Details on the numerical simulations of the system evolution can be found in Appendix \ref{Appendix1}, whereas PPO algorithm informations, hyperparameters and Neural Network architectures can be found in Appendix \ref{Appendix2}.

\subsection{Optimal design}
\label{sec:R_opt}
We start by considering the case where the target site is the sink and the Agent makes no errors in placing the particles in the desired locations.

As a first test, we applied the approach described in Section \ref{sec:RL} when $J=\Delta E$, for which the excitation transfer is already high, as it provides $p_{sink}(T)\approx0.86$. With $N_{cells}=11$, we find $p_{sink}(T)>0.99$ after applying our RL approach. The optimum consists in placing a third particle right half-way between $A$ and $B$, as it might have been guessed.
We also noticed that filling all the cells with particles yields $p_{sink}(T)\approx0.97$, making the RL solution more effective with noticeably less resources.
For $J/\Delta E=0.05$, the best configuration for $N_{cells}=11$ turns out to be 
\begin{equation}
    S_{opt} = 10101010101,
\end{equation}
where $0$ and $1$ stand for the absence or presence of a particle in a cell (including $A$ and $B$). We thus obtain a configuration of equally spaced particles, which allows a population transfer to the sink of $p_{sink}(T)\approx0.98$. 
However, these features are not general. For instance, by increasing the number of cells to $N_{cells}=21$ allows us to achieve a population transfer of $p_{sink}(T)>0.99$ through the asymmetric configuration
\begin{equation}\label{opt_sink_noerrors}
    S_{opt} = 100000100010010010001.
\end{equation}
It is interesting to notice that we only need to add $4$ unevenly distributed particles to realize the ideal chain configuration, thus raising the performance of excitation transfer -- in the given time -- from negligible to nearly perfect. This result is even more noticeable when compared to the naive decision to fill all the available cells with particles, which corresponds to  $p_{sink}(T)\approx0.97$.

\subsection{Addressing errors and disorder}
\label{sec:R_err}

\begin{figure}
\centering\includegraphics[width=\columnwidth]{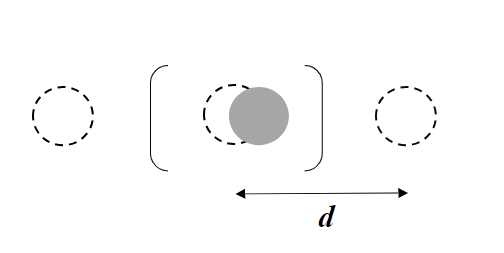}
	\caption{Error made by the agent in the positioning operation. The particle is not found exactly at the desired location. However we assume that the error is not large enough to make the assignation of the particle to a given spatial cell undecidable.}\label{Figure:errors}
\end{figure}

So far we have consider the ideal case where no source of disorder is present. Such an assumption can be relaxed to address the effects of imperfections in the performance of our protocol. We thus consider the case in which the positioning of the particles across the chain is affected by static disorder. 
We can thus associate an uncertainty $\delta x_j$ to the position of the $j^\text{th}$ element of a chain as determined by the action of the Agent, who will allocate the particle at position $x_j\pm \delta x_j$. We further assume that
\begin{equation}\label{eq:error}
   \delta x_j < d/2 \, , 
\end{equation}
where $d = d_{AB}/(N_{cells}-1)$ is the distance between two adjacent spatial cells, so that we can still assign the particle to cell $j$ (as the latter is the closest cell). This implies that the cells are still distinguishable [cf. \Cref{Figure:errors}]. These errors will affect the expected dynamics, but will not change the formulation of the MDP. The observation of the Agent can still be expressed as a string of dichotomic variables (being either $0$ or $1$).

We first considered uniformly distributed errors, i.e. $\delta x_j \in \mathcal{U} ( [0, r {d}/{2}])$, with $0 < r < 1$, where $\mathcal{U} ( [\alpha, \beta]) $ denotes the uniform distribution over the interval $ [\alpha, \beta]$. For $r=0.1$, the optimal chain configuration in \Cref{opt_sink_noerrors} already yields $p_{sink}(T)>0.99$. This is no longer true if we increase the maximum error. For $r=0.25$, the optimal configuration is
\begin{equation}\label{opt_25}
    S_{opt} = 100010001000100010001,
\end{equation}
whose Hamming distance with the string in Eq.~\eqref{opt_sink_noerrors} is $6$. For $r=0.5$, we have
\begin{equation}\label{opt_50}
    S_{opt} = 100001000010000100001,
\end{equation}
which has an Hamming distance $5$ with the string in Eq.~\eqref{opt_sink_noerrors} and $7$ with the string in Eq.~\eqref{opt_25}.

We also considered the case of normally distributed errors, i.e. $\delta x_j \in \mathcal{N}(0,\sigma)$, where $\mathcal{N}(\mu,\sigma)$ denotes the normal distribution with mean $\mu$ and standard deviation $\sigma$. Note that \Cref{eq:error} sets a bound on the maximum extracted values, while we consider $\sigma$ such that $5 \sigma = d/2$. We found the optimal configuration to be given by \Cref{opt_25}. All results are reported in \Cref{tab:errors}, which show that, for different error distributions, we are always able to reach higher population transfers than those obtained by the mere application of \Cref{opt_sink_noerrors}. Interestingly, we gather numerical evidence that, as opposed to case of small or no errors, the best chain configurations for large error are given by equally spaced particles.  Moreover, we found that applying either \Cref{opt_25} or (\ref{opt_50}) to the case of low or no errors leads to worse results than those achieved through \Cref{opt_sink_noerrors}, making the equally spaced particles solutions characteristic of the moderate-high disorder scenario.

\begin{table}[t]
\begin{tabular}{c | c | c |c |c } 
 $\delta x_j$ & Chain & $\langle p^{opt} \rangle$ & $\langle p^{\delta x_j=0} \rangle$ & $\langle p^{filled} \rangle$\\ [0.5ex] 
 \hline\hline
 $\mathcal{U} ([0,{d}/{20}] )$ & Eq.~\eqref{opt_sink_noerrors} & $0.998$ & $0.998$ & $0.796$\\ 
 \hline
 $\mathcal{U} ([0,{d}{/8}] )$ & Eq.~\eqref{opt_25} & $0.995$ & $0.989$ & $0.300$\\
 \hline
 $\mathcal{U} ([0,{d}/{4}] )$ & Eq.~\eqref{opt_50} & $0.988$ & $0.943$ & $0.045$\\
 \hline
 $\mathcal{N}(0, {d}/{10})$ & Eq.~\eqref{opt_25} & $0.979$ & $0.972$ & $0.188$\\ [0.5ex] 
\end{tabular}\caption{From the left: error probability distribution, optimal chain found with the RL approach, average $p_{sink}(T)$ over $5000$ random chain extractions using the corresponding optimal chain and error distribution, average $p_{sink}(T)$ over $5000$ random chain extractions using the chain given in \Cref{opt_sink_noerrors} perturbed by the corresponding error distribution and average $p_{sink}(T)$ over $100$ random chain extractions when all the cells are filled, perturbed by the corresponding error distribution.}\label{tab:errors}
\end{table}

However, we have not leveraged the full potential of RL so far, as we were mostly interested in the optimal chain discovered through policy learning, rather than the agent's policy itself. This change of perspective leaves room for further improvement of the excitation transfer, provided that we are not interested in a single chain configuration that maximizes the average transfer. Alternatively, we might be interested in finding a way to optimize each chain configuration adaptively, taking into account the specific disorder without measuring the errors in the particle positioning. To this end, we can notice that, even if we do not measure the positions of the particles, we already gain some information on the errors made during the particles positioning operations. For the Agent to receive its reward,  we need to measure at each step the target population $p_{sink}(T)$, which is itself implicitly affected by the positions of the particles in the cells. We replace the $1$'s in the string  describing the environment state with a function of $p_{sink}(T)$ for the chain configurations obtained when adding the particles. Therefore, the Agent can take different actions depending on such values, which in turn depend on the current information available on the disorder. In particular, at each step, after adding a particle to the chain, we change the value of the corresponding cell in the string that represents the environment state from $0$ to
\begin{equation}
   \eta = \frac{1+p_{sink}(T)}{2}\in[1/2,1].
\end{equation}
This choice yields the desired information more effectively, as it is still well separated from the case in which a cell in empty.

We apply this approach to the case of $\delta x_j \in \mathcal{N}(0,\sigma)$, for which the previous method was less effective: we obtained $\langle p_{sink}(T) \rangle = 0.979$ over $5000$ simulations. By contrast, using the whole Agent's policy learned with the new version of the environment state, we were able to visibly improve the excitation transfer, obtaining $\langle p_{sink}(T) \rangle = 0.987$ over $5000$ simulations, with an average of $5$ particles in the chain.

\subsection{Unitary case}
\label{sec:R_uni}

\begin{figure}
\centering
	\textbf{(a)}\\
 \centering\includegraphics[width=\columnwidth]{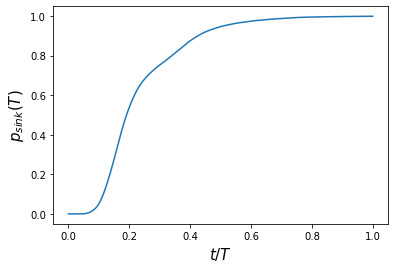}\\
 \textbf{(b)}\\
 \includegraphics[width=\columnwidth]{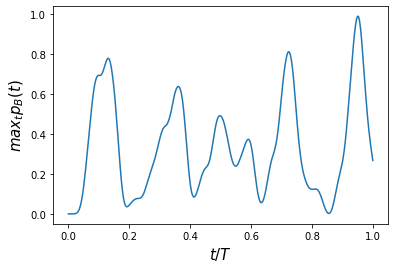}
	\caption{\textbf{(a)} Final excitation transferred to the sink 
 and \textbf{(b)} maximum excitation transferred to B within the evolution time $T$ as function of time when the particle chains are given by the optima found during the RL agent's learning (\Cref{opt_sink_noerrors} and \Cref{opt_nosink}, respectively).}\label{Figure:poptransfer}
\end{figure}

Our discussion has been hitherto focused only on the the case in which the excitation in the system is irreversibly transferred to a sink via spontaneous emission. This results into a monotonic behaviour of the target (i.e. sink) excitation over time, as shown in \Cref{Figure:poptransfer} \textbf{(a)}. Hence, it makes sense to frame the problem as the optimization of the target population at a time $T$. However, this scenario might be of limited interest for applications, especially for quantum communication purposes, as the irreversibility of the process causes the system to lose coherence. To circumvent the coherence loss, we can remove the sink and directly consider the population transferred to $B$. This choice results into a different time behaviour of the target excitation: it does not increase monotonically in time, as shown in \Cref{Figure:poptransfer} \textbf{(b)}. Therefore, instead of looking at the excitation transferred at a time $T$, it is better to optimize the maximal excitation transfer within a time interval $T$, i.e. $\max_{t\in [0,T]} p_B(t)$.

This requires to sample the dynamics at different instants of time. In our simulations we considered $n_T=20$ equally spaced points over the time interval $[0,T]$ to calculate the maximum population transfer, hence the reward $R_{i+1}$ at each step. Besides the change in the definition of the Agent's reward, the MDP is identical to the scenario where the target is the sink. For this case, we assume no errors are made in the particle positioning operation as in \Cref{sec:R_opt}. The optimal chain configuration found during learning is
\begin{equation}\label{opt_nosink}
    S_{opt} = 100010111010111010001,
\end{equation}
for which $\max_{t\in [0,T]} p_B(t)>0.99$.

Notice that in this instance, despite the particles not being all equally spaced, the chain is symmetrical; furthermore, compared to the sink scenario, we need a larger number of particles  to realize almost perfect excitation transfer. 
In this case, filling all the cells with particles yields $\max_{t\in [0,T]} p_B(t)\approx0.3$. We also noticed that, using \Cref{opt_nosink} for the sink-target case, we could still achieve high final excitation transfer ($0.991$ compared to the optimum $0.998$). Conversely, using \Cref{opt_sink_noerrors} in the no-sink scenario, we found the maximum excitation transfer to be low, i.e. $\max_{t\in [0,T]} p_B(t)\approx0.5$.

\section{Conclusions}
\label{sec:conclusions}
We have developed a spatial, bottom-up, RL-based approach to the design of particle chains for optimal excitation transfer. We studied the effectiveness of such approach under different conditions. In particular, we considered two different scenarios, i.e. with or without a sink attached to the chain. In the former case, where we 
we are not interested in preserving coherence, we consider our target to be a sink where excitation is irreversibly transferred from the end of the chain. In the latter case, instead, since we want to avoid coherence loss, our target is the last particle of the chain. We also tested our approach in the presence of agent's errors, adapting our technique to minimize their effects when we build the chain.

We were able to achieve extremely high excitation transfer in all scenarios, resulting into different particle chain design for each case. Our solutions exhibit some interesting properties. In particular, we found an optimal asymmetrical chain with a smaller number of particles in the sink scenario, while the optimal chain found in the no-sink case is symmetrical, presents a peculiar structure, and it is made of a larger number of particles. In the presence of moderate or large agent's errors (which we studied only in the sink case), we found that, to maximize the average excitation transfer, the optimal chains are composed of equally spaced particles, where the spacing dependent of the amount of errors. If we are willing to renounce to a single chain design for all disorder configurations from the error distribution, the full potential of RL can be deployed: we can adaptively build the optimal chain without making additional measurements, further improving the excitation transfer in this case.

Our approach presents multiple advantages compared to other techniques. In particular, the spatial dependency of the couplings makes it closer to realistic physical problems and allows us to go beyond some of the usual approximations (e.g. the nearest-neighbour approximation), without rendering the optimization problem extremely complex. Furthermore, when we rely on RL, we do not need to fix the number of particles beforehand, though  we can limit such number, hence the resources used to build the optimal chain.

Finally, the methodology presented can easily be extended beyond the scenario considered here, as long as the Hamiltonian controlling the interaction between the particles depends on their relative position in space. In principle, one can introduce some changes without substantially affecting the formulation of the problem in terms of MDP. For instance, we could change the specific form of the interaction, we could allow the particles' local energies depend on their positions, we could add some environmental effects, or go beyond the one-excitation approximation. More complex scenarios can be then addressed with the same technique, as long as we are able to simulate the system dynamics or measure the amount of transfer. Its effectiveness for more complex cases is yet to be ascertained, but, given the widespread success of RL for complex tasks, it is reasonable to believe that this approach might still work. It would then be relevant to apply it to a more realistic model of a technologically relevant quantum system, maybe in a real experimental setting.
It would also be interesting to extend the use or RL for quantum system design to solve different physical problems.

\acknowledgements
AI gratefully acknowledges the financial support of The
Faculty of Science and Technology at Aarhus University
through a Sabbatical scholarship and the hospitality of the
Quantum Technology group, the Centre for Quantum Materials and Technologies, and the School of
Mathematics and Physics, during his stay at Queen's University Belfast. MP acknowledges  support from 
the Horizon Europe EIC-Pathfinder project QuCoM (101046973), 
the Leverhulme Trust Research Project Grant UltraQuTe (grant RGP-2018-266), the Royal Society Wolfson Fellowship (RSWF/R3/183013), the UK EPSRC (EP/T028424/1), and the Department for the Economy Northern Ireland under the US-Ireland R\&D Partnership Programme.

\bibliography{biblio.bib}

\appendix
\section{Numerical Simulations}\label{Appendix1}
In order to solve the system dynamics, we first vectorise Eq. \ref{eq:ME} \cite{amshallem2015approaches}. This transforms the density matrix as
\begin{equation}\label{eq:r}
    \rho\rightarrow\Vec{r}=(\rho_{00},\rho_{01},...,\rho_{0 K}, \rho_{10},...,\rho_{K K})
\end{equation}
with $K = N+1$ in the presence of the sink, and $K = N$ in the unitary case. Here,  $N$ is the current number of particles in the chain. The unitary part of the master equation becomes
\begin{equation}\label{eq:TH}
    [H,\rho] \rightarrow \mathcal{L}_U\Vec{r} \equiv (I\otimes H-H^T\otimes I)\Vec{r},
\end{equation}
while the dissipative one transforms as
\begin{equation}
\begin{aligned}
\label{eq:Lindbladian_vec}
    &L\rho L^{\dagger}-\frac{1}{2}\{L^{\dagger} L, \rho\}\rightarrow\\
    &\mathcal{L_{D}}\Vec{r}= \Bigg[(L^\dagger)^T\otimes L-\frac{1}{2}(I\otimes {L}^\dagger {L}+({L}^\dagger {L})^T\otimes I)\Bigg]\Vec{r}.
\end{aligned}
\end{equation}
By defining $\mathcal{L} = \mathcal{L_{D}}+\mathcal{L_{U}}$, we obtain
 $   \dot{\vec{r}} = \mathcal{L} [ \vec{r}  (t) ]$,
hence the state of the system at a time $t$ reads 
$    \Vec{r}(t)=e^{-i t \mathcal{L}} \; \Vec{r}(0)$,
where $\Vec{r}(0) = (1, 0, ..., 0)$.
To calculate the excitation transfer at a time $t$, we project $\Vec{r}(t)$ into the target $r_{target}^t=\Vec{r}(t)\cdot\Vec{r}_{target}$. Note that $\Vec{r}_{target}$ is either the $N+1$ dimensional vector $\Vec{r}_{S} = (1,0,...,0)$ in the sink case, or the $N$ dimensional vector $\Vec{r}_{B} = (1,0,...,0)$ in the no-sink case. Then, in the former case, we simply have $p_{sink}(T)=r_{target}^T$ while in the latter $\max_{t\in [0,T]} p_B(t) \approx \max(\Vec{r}(0), ..., r_{B}^{t_n}, r_{B}^{t_{n+1}}, ..., r_{B}^{T})$, where we have divided the time interval $[0,T]$ in $n_T$ equally spaced points, as explained in Section \ref{sec:R_uni}. 

We performed all our numerical calculations using Python, in particular the modules NumPy \cite{harris2020array} and SciPy \cite{2020SciPy-NMeth}.

\section{Algorithm hyperparameters and Neural Network architectures}\label{Appendix2}
Throughout this work we used the clipped PPO algorithm described in Ref.~\cite{schulman2017proximal} with clipping parameter $\epsilon = 0.2$, $100$ agents and $4$ epochs of learning with minibatch size $128$ in all cases except for the uniform error distributions in Section \ref{sec:R_err}, where the minibatch size was $64$. The number of episodes considered was always $<1500$. We fixed the discount factor $\gamma=0.99$ and the parameter $\lambda = 0.95$ for the generalized advantage estimation, while $\lambda = 0.98$ in \Cref{sec:R_uni}.

We used two separate Neural Networks for the Actor and the Critic. All hidden layers have a ReLu activation function and the output layer for the Critic has a linear activation function while the ouptut for the Actor is given by a softmax activation function. The Actor has $2$ hidden layers of $128$ neurons in all cases except for the adaptive error case described at the end of Section \ref{sec:R_err}, where the hidden layers are $3$. The Critic has $2$ layers of $64$ Neurons in all no-error cases and $3$ hidden layers of $128$ neurons in all other cases except the adaptive case, where the hidden layers are $4$. The optimizations were performed using Adam \cite{kingma2017adam} with different learning rates $lr_A$ and $lr_C$ for the Actor and the Critic, respectively [cf.~\Cref{tab:values} for the values being used in each Section of the paper]. 

\begin{table}[b]
\begin{tabular}{c | c | c} 
Section & $lr_A$ & $lr_C$ \\ 
 \hline
\hline 
 \Cref{sec:R_opt}&$3\times10^{-4}$&$5\times10^{-4}$\\
 \Cref{sec:R_err}&$1\times10^{-4}$&$1\times10^{-4}$\\
 \Cref{sec:R_uni}&$8\times10^{-5}$&$1\times10^{-4}$\\
 \hline\hline
\end{tabular}\caption{Values of the learning rates used for Actor and Critic in the various case studies reported in the paper. }\label{tab:values}
\end{table}

All Neural Networks and their parameters optimization were implemented using Tensorflow \cite{tensorflow2015-whitepaper} and Keras \cite{chollet2015keras}.

\end{document}